\documentclass[letterpaper,aps,prb,twocolumn,showpacs]{revtex4}
\usepackage{graphicx}
\usepackage{amsmath}
\usepackage{epstopdf}
\usepackage{amsfonts}
\usepackage{amssymb}

\begin{document}
\title{Boundary spin Hall effect in a two-dimensional semiconductor system with Rashba spin-orbit coupling}
\author{Yaroslav Tserkovnyak}
\affiliation{Lyman Laboratory of Physics, Harvard University, Cambridge, Massachusetts 02138, USA}
\affiliation{Department of Physics and Astronomy, University of California, Los Angeles, California 90095, USA}
\author{Bertrand I. Halperin}
\affiliation{Lyman Laboratory of Physics, Harvard University, Cambridge, Massachusetts 02138, USA}
\author{Alexey A. Kovalev}
\affiliation{Department of Physics, Texas A\&M University, College Station, Texas 77843, USA}
\author{Arne Brataas}
\affiliation{Department of Physics, Norwegian University of Science and Technology, NO-7491 Trondheim, Norway}
\affiliation{Centre for Advanced Study at the Norwegian Academy of Science and Letters, Drammensveien 78, NO-0271 Oslo, Norway}

\begin{abstract}
We derive boundary conditions for the coupled spin-charge diffusion equations at a transmitting interface between two-dimensional electron systems with different strengths of the Rashba spin-orbit (SO) coupling $\alpha$, and an electric field parallel to the interface. We consider the limit where the spin-diffusion length $l_s$ is long compared to the electron mean free path $l$, and assume that $\alpha$ changes discontinuously on the scale of $l_s$. We find that the spin density is also discontinuous on the scale of $l_s$. In the case  where the electron mobility is constant across the interface, this leads to the complete suppression of the expected spin injection from a region with $\alpha\neq0$ into a non-SO region with $\alpha=0$.
\end{abstract}

\date{\today}
\maketitle


\section{Introduction}

Developing robust mechanisms for spin generation and detection is a central objective in the rapidly growing field of spintronics.\cite{wolfSCI01} There are many exciting venues for basic physics and nanotechnology to exploit injected spins in various systems, in both the semiclassical transport regime and the quantum regime where spins can encode entangled quantum bits. Since the early days of spintronics,\cite{dattaAPL90} however, exciting theoretical predictions often seemed to be plagued by experimental ambiguities. A notable exception is the field of metallic magnetoelectronics,\cite{tserkovRMP05} where many of key experimental findings have been well understood and the theoretical predictions in turn were instrumental in pushing the experimental frontier. Semiconductor spintronics which possesses much richer phenomenology and appears more desirable for technological applications, however, has caused more problems for theorists and experimentalists alike.

In systems with intrinsic spin-orbit (SO) coupling, the spin degrees of freedom are intricately entangled with the orbital motion, and even the concept of spin current (defined as a symmetrized product of spin and velocity operators) has resisted the thorough qualitative understanding.\cite{engelCM06} There is an opinion that for many experimental implications, it is more relevant to calculate spin densities rather than spin currents. This is a principal motivation of this paper. The spin-density generation and dynamics in diffuse bulk semiconductors are conveniently described by the semiclassical diffusion equation which is usually derived in the limit of weak SO coupling and dilute disorder,\cite{mishchenkoPRL04,burkovPRB04,bleibaumPRB06sd,adagideliPRL05, malshukovPRL05sh} $\Delta,\tau^{-1}\ll E_F$, where $\Delta$ is the characteristic SO splitting, $\tau^{-1}$ is the impurity scattering rate, and $E_F$ is the Fermi energy (setting $\hbar=1$ here and henceforth). In order to calculate the nonequilibrium spin accumulation generated by the spin Hall effect at a Hall-bar edge or at a boundary between two different conductors, the diffusion equation has to be supplemented by the appropriate boundary conditions (BC) for spin and charge densities. The latter have been the source of a vigorous discussion in recent literature.\cite{bleibaumPRB06sd,adagideliPRL05, malshukovPRL05sh,galitskiPRB06,bleibaumPRB06bc,rashbaPE06} In addition, the boundary problem with SO interactions was previously discussed in the context of spin-polarized ballistic beam reflection and refraction\cite{khodasPRL04} and in the regime of Friedel-like spin-density oscillations on the scale of the Fermi wavelength at sharp boundaries.\cite{reynosoPRB06}

In this paper, we develop a general approach for systematically deriving semiclassical BC for diffuse spin transport with weak intrinsic SO interaction.  We assume that $E_F\tau\gg1$, but we  work in the ``SO dirty limit," where $\Delta\tau \ll1$ and the D'yakonov-Perel spin-diffusion length $l_s$ is much larger than the electron mean free path $l$. (This limit is, in fact, necessary for the diffusion equation itself to be meaningful near a boundary.) We use a Keldysh kinetic equation approach,\cite{rammerRMP86} similar in spirit to the one used in Ref.~\onlinecite{mishchenkoPRL04} for deriving the spin-diffusion equation. Our method reveals the physical meaning of the BC, which possesses an interesting correspondence with the classical Hall physics. We believe that our findings resolve some discrepancies existing in the literature.

\section{Kinetic equation approach}

Of central importance as a convenient case study as well as an experimentally relevant model is the noninteracting Rashba Hamiltonian which describes the simplest two-dimensional system with intrinsic SO coupling:\cite{rashbaPE06}
\begin{align}
H&=\frac{\mathbf{\tilde{p}}^2}{2m} +U(\mathbf{r})-E_F+H_{\rm{SO}}\,,\nonumber\\
H_{\rm SO}&=-\frac{1}{2}\left\{\alpha(\mathbf{r}),\tilde{p}_x\right\}\hat{\sigma}_y+\frac{1}{2}\left\{\alpha^\prime(\mathbf{r}),\tilde{p}_y\right\}\hat{\sigma}_x\,,\label{Hsoxy}
\end{align}
where $\mathbf{\tilde{p}}=\mathbf{p}-e\mathbf{A}$, $\mathbf{p}=-i\boldsymbol{\nabla}$ is the canonical momentum, $e$ the electron charge, $\mathbf{A}$ a vector potential, $m$ the effective electron mass, $\boldsymbol{\hat{\sigma}}$ a vector of Pauli matrices, $U$ the potential due to disorder and external gates, and  $\alpha$ is the Rashba SO interaction parameter, which here depends on the position. The second coupling constant $\alpha^\prime$ is equal to $\alpha$ for pure Rashba coupling, but  the situation $\alpha^\prime \neq \alpha$ applies, e.g., to structures on a [001] surface of GaAs, when linear Dresselhaus coupling is  present, if the $x$ axis is chosen along the (110) crystal direction. Note that we are disregarding the SO coupling due to the lateral potential $U$, and the electrons are moving in the $xy$ plane.

\begin{figure}
\includegraphics[width=\linewidth,clip=]{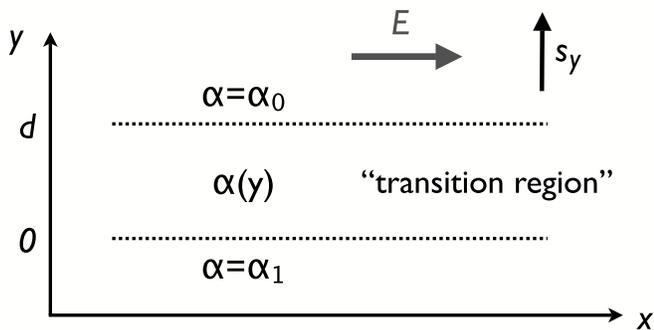}
\caption{Simple example: A two-dimensional electron gas in the $xy$ plane with a uniform electric field $\bf{E}$ in the $x$ direction, and an inhomogeneous Rashba coupling $\alpha(y)$, which varies between $\alpha=\alpha_1$, for $y<0$, and $\alpha=\alpha_0$, for $y>d$. If the in-plane spin polarization $s_y$, which occurs in the region $y>d$ spills over into the normal region $y<0$, there will be a nonzero value of $s_z$ near the boundary, which can be measured optically. We find, however, that if the electron mobility is independent of $y$,  there is no spillover of $s_y$, and $s_z=0$.}
\label{sc}
\end{figure}

Although our approach is general and can be applied to other SO Hamiltonians and different heterostructures,
we will focus on the BC in the Hall configuration of Fig.~\ref{sc}. Specifically, we assume a uniform electric field $\mathbf{E}=-\partial_t\mathbf{A}$ parallel to the $x$ axis, with a reflecting wall at some negative $y$ blocking any transverse charge currents. The two-dimensional electron gas is translationally invariant along the $x$ axis, apart from the disorder potential, but the system parameters depend explicitly on $y$.  We assume that the Rashba parameter $\alpha$ has different constant values $\alpha_1$ and $\alpha_0$ for $y<0$ and $y>d$, with a ``transition region'' $0<y<d$. For reasons that will become clear later, we choose $d$ to be long on the scale of the mean-free path $l=v_F\tau$ ($v_F=\sqrt{2E_F/m}$ is the Fermi velocity), but short on the scale of the D'yakonov-Perel spin-diffusion length $l_s=l/\Delta_0\tau$ (with $\Delta_0=\alpha_0p_F$ and $p_F=mv_F$), so that the spins do not relax while traversing the transition region. Note that by  definition, $l_s=\sqrt{D\tau_s}$, where $D$ is the diffusion constant and $\tau_s$ a characteristic spin-relaxation time. A geometry of this type can be produced by a charged gate covering one half of the 2D system, say the region $y<0$, in which case the transition length $d$ will be determined by the setback distance of the gate. In such a system, not only $\alpha$, but also the potential $U$, the electron density $n$, and the electron mobility $e \tau/ m$ will generally depend on $y$. However, for pedagogical purposes, we shall first consider a simplified model, where we assume that only $\alpha$ depends on $y$.  Furthermore, we assume pure Rashba coupling, and we take $\alpha_1=0$.  

The kinetic equation with $y$-dependent $\alpha$ can be derived using the approach already discussed in Ref.~\onlinecite{mishchenkoPRL04}, with the result
\begin{widetext}
\begin{equation}
\partial_t\hat{g}+\frac{1}{2}\left\{\frac{\mathbf{p}}{m}+\alpha\boldsymbol{\hat{\eta}},(\boldsymbol{\nabla}+e\mathbf{E}\partial_\epsilon)\hat{g}\right\}+i\alpha\mathbf{p}\cdot\left[\boldsymbol{\hat{\eta}},\hat{g}\right]-\frac{\partial_y\alpha}{2}\left\{\boldsymbol{\hat{\eta}}\cdot\mathbf{p},\partial_{p_y}\hat{g}\right\}=\frac{1}{2\tau}\left(\left\{\hat{\rho},\hat{A}\right\}-\left\{\hat{g},\frac{1}{m}\int\frac{dp^2}{(2\pi)^2}\hat{A}\right\}\right)
\label{ke}
\end{equation}
\end{widetext}
containing an additional term proportional to $\partial_y\alpha$, which is at the heart of our discussion. Here, $\boldsymbol{\hat{\eta}}=\mathbf{z}\times\boldsymbol{\hat{\sigma}}$, with $\mathbf{z}$  the unit vector along the $z$ axis, and
\begin{equation}
\hat{\rho}(\epsilon)=\frac{1}{m}\int\frac{d^2p}{(2\pi)^2}\hat{g}(\mathbf{p},\epsilon)\,.
\end{equation}
The second term on the left-hand side of Eq.~(\ref{ke}) is the drift with spin-dependent velocity, the third term is the spin precession in the Rashba field, and the fourth term is the spin-dependent acceleration in the $y$-dependent Rashba field. The right-hand side is the collision integral in the self-consistent Born approximation due to the weak and isotropic scalar disorder. $\hat{A}=i(\hat{G}^R-\hat{G}^A)$ is the spectral function in terms of the retarded and advanced Green's functions $\hat{G}^{R,A}$, which is given by
\begin{equation}
\hat{A}(\mathbf{p},\epsilon)=\frac{A_\uparrow+A_\downarrow}{2}+\hat{\eta}_{\theta}\frac{A_\uparrow-A_\downarrow}{2}\,.
\end{equation}
Here,
\begin{equation}
A_s(p,\epsilon)=\frac{1/\tau}{(\epsilon-\epsilon_{ps})^2+(1/2\tau)^2}\,,
\end{equation}
$\hat{\eta}_{\theta}=\boldsymbol{\hat{\eta}}\cdot\mathbf{p}/p=\sin\theta~\hat{\sigma}_x-\cos\theta~\hat{\sigma}_y$, where $\theta$ is the angle of $\mathbf{p}$ with respect to the $\mathbf{x}$ axis, and
\begin{equation}
\epsilon_{ps}=\frac{p^2}{2m}-E_{F}+s\alpha|p|\,,
\end{equation}
with $s=\pm$ corresponding to $\uparrow,\downarrow$. We have neglected the real part of the Green's functions $\hat{G}^{R,A}$ in the collision integral of the kinetic equation (\ref{ke}), since it only results in a small correction of order $(\tau E_F)^{-1}$ to the Rashba precession term.

After solving the kinetic equation (\ref{ke}), we can find the spin density $\mathbf{s}(\mathbf{r},t)$ in terms of the Keldysh Green's function $\hat{g}(\mathbf{r},t;\mathbf{p},\epsilon)$ (in the Wigner representation),\cite{rammerRMP86}
\begin{equation}
\mathbf{s}(\mathbf{r},t)=\frac{1}{2}\int\frac{dp^2d\epsilon}{(2\pi)^3}\mbox{Tr}\left[\boldsymbol{\hat{\sigma}}\hat{g}(\mathbf{r},t;\mathbf{p},\epsilon)\right]\,.
\label{s}
\end{equation}
In equilibrium,\cite{rammerRMP86}
\begin{equation}
\hat{g}_0(\mathbf{r},t;\mathbf{p},\epsilon)=-\frac{\tanh(\beta\epsilon/2)}{2}\hat{A}(\mathbf{r},t;\mathbf{p},\epsilon)\,,
\end{equation}
where $\beta=1/k_BT$ is the inverse temperature, and we will be interested in the $T\to0$ limit in the following. The kinetic equation can be simplified after defining the distribution function $\hat{f}(\epsilon,\theta)$ at given position and time:
\begin{equation}
\hat{g}(\epsilon,\mathbf{p})=\frac{1}{2}\left\{\hat{A}(\epsilon,\mathbf{p}),\hat{f}(\epsilon,\theta)\right\}\,,
\label{f}
\end{equation}
where we assume that at given energy $\epsilon$ and momentum direction $\theta$, the Keldysh Green's function $\hat{g}$ has the momentum profile determined by the spectral function, i.e., it can be expressed as $\hat{C}_\uparrow(\epsilon,\theta)A_\uparrow(p,\epsilon)+\hat{C}_\downarrow(\epsilon,\theta)A_\downarrow(p,\epsilon)$ in terms of some Hermitian matrix coefficients $\hat{C}_{\uparrow,\downarrow}$. Note that in equilibrium, $\hat{f}(\epsilon,\theta)\equiv-\tanh(\beta\epsilon/2)/2$. Ansatz~(\ref{f}) allows us to integrate Eq.~(\ref{ke}) over the momentum ($\int pdp/2\pi$) at fixed direction $\theta$ and energy $\epsilon$. Then we can integrate over the energy $\epsilon$ and look for a steady-state solution. To this end, we have to solve the following transport equation for the distribution function:
\begin{widetext}
\begin{align}
\hat{s}_E+\sin\theta~\partial_y\hat{f}+\frac{\gamma}{2}\left\{\cos2\theta~\hat{\sigma}_x+\sin2\theta~\hat{\sigma}_y,\partial_y\hat{f}\right\}+i\gamma p_F\left[\hat{\eta}_\theta,\hat{f}\right]&-\frac{\partial_y\gamma}{2}\cos\theta\left\{\hat{\eta}_\theta,\partial_\theta\hat{f}\right\}\nonumber\\
&=\frac{1}{2l}\left(\left\{\left\langle\hat{f}\right\rangle_\theta-\hat{f},1-\gamma\hat{\eta}_\theta\right\}-\gamma\left\langle\left\{\hat{\eta}_\theta,\hat{f}\right\}\right\rangle_\theta\right)\,,
\label{fe}
\end{align}
\end{widetext}
where $\hat{f}(\theta)=\int d\epsilon\hat{f}(\epsilon,\theta)$, $\langle\rangle_\theta$ denotes angular averaging and $\gamma(y)=\alpha(y)/v_F$. $\hat{s}_E$ is the source term proportional to the electric field,
\begin{equation}
\hat{s}_E=eE\left[-\cos\theta+\gamma(\sin2\theta~\hat{\sigma}_x-\cos2\theta~\hat{\sigma}_y)\right]\,,
\label{S}
\end{equation}
and we have to supply Eq.~(\ref{fe}) with the ``initial condition,"
\begin{equation}
y=0:~~~\hat{f}(\theta)=\hat{f}_0+eEl\cos\theta\,.
\label{bc}
\end{equation}
Here, $\hat{f}_0$ is an isotropic spin and charge imbalance (assumed to be linear in $\gamma$ as it vanishes for $\gamma=0$), which can in general be induced by the electric field in the bulk and leak towards the normal boundary. $\hat{f}_0$ is assumed to be angle $\theta$ independent, since any nonisotropic component of $\hat{f}_0$ decays on the scale of the mean free path at $y<0$, if we take $\alpha_{1}=0$. The second term in Eq.~(\ref{bc}) is the usual drift along the $x$ axis. We can now integrate Eq.~(\ref{fe}) using Eqs.~(\ref{S}) and (\ref{bc}). For the BC across the transition region $0<y<d$ which complements the diffusion equation (that is second order in spatial derivatives), we need to derive the expressions for the spin density $\mathbf{s}$ and $\partial_y\mathbf{s}$ at $y=d$, i.e., after $\gamma$ is fully turned on, for a given spin density at $y=0$ (which can then be determined self-consistently after solving the diffusion equation). Working to the lowest nontrivial order in $\gamma$, we have to calculate $\mathbf{s}$ to the first order and $\partial_y\mathbf{s}$ to the second order, as will be explained below.

\section{Boundary spin accumulation}

To linear order in $\gamma$, Eq.~(\ref{fe}) simplifies tremendously to
\begin{equation}
\hat{s}^\prime_E+\sin\theta~\partial_y\hat{f}+eEl\frac{\partial_y\gamma}{2}\sin2\theta~\hat{\eta}_\theta=\frac{1}{l}\left(\left\langle\hat{f}\right\rangle_\theta-\hat{f}\right)\,,
\label{se}
\end{equation}
where
\begin{equation}
\hat{s}^{\prime}_E=eE\left[-\cos\theta+\frac{\gamma}{2}(\sin2\theta~\hat{\sigma}_x-\cos2\theta~\hat{\sigma}_y)\right]\,.
\end{equation}
Equation (\ref{se}) was obtained by inserting the normal solution $\hat{f}_N=eEl\cos\theta$ into terms in Eq.~(\ref{fe}) which have prefactors linear in $\gamma$. Averaging Eq.~(\ref{se}) and also Eq.~(\ref{se}) multiplied by $\sin\theta$ over $\theta$, we get
\begin{equation}
\left\langle\hat{f}(y)\right\rangle_\theta=\hat{f}_0+eEl\frac{\gamma(y)}{4}\hat{\sigma}_y+\left\langle\cos2\theta~\hat{f}(y)\right\rangle_\theta\,.
\label{fy}
\end{equation}
We can evaluate the last term in this equation by noticing that if we fix $\langle\hat{f}\rangle_\theta$ and $\partial_y\gamma$ in Eq.~(\ref{se}), then impurity scattering equilibrates (on the scale of the mean free path) the distribution to
\begin{equation}
\hat{f}(y)\approx\left\langle\hat{f}(y)\right\rangle_\theta-l\left(\hat{s}^{\prime}_E+eEl\frac{\partial_y\gamma}{2}\sin2\theta~\hat{\eta}_\theta\right)\,,
\end{equation}
where the last term should actually also be neglected in the spirit of the approximation (which disregards corrections of order $l/d\ll1$ with respect to the leading term linear in $\gamma$). Combining this with Eq.~(\ref{fy}), we get
\begin{equation}
\hat{f}(y)=\hat{f}_0+eEl\cos\theta\left[1-\gamma(y)\hat{\eta}_\theta\right]\,.
\label{fy0}
\end{equation}
This gives for the spin density (for the case $\alpha_1=0$)
\begin{equation}
\mathbf{s}(y)=\mathbf{s}(0)+\nu eE\tau\alpha(y)\mathbf{y}\,,
\label{sd}
\end{equation}
where $\nu=m/2\pi$ is the density of states per spin. This is just the position-dependent bulk result\cite{mishchenkoPRL04} offset by $\mathbf{s}(0)$. The charge density $n(y)$ is uniform. It may appear quite surprising (although we will provide a physical explanation in the next section) that the spin density follows the Rashba parameter profile which changes rapidly on the scale of the spin-diffusion length $l_s$.

Inserting the distribution function (\ref{fy0}) into the terms which already have prefactors linear in $\gamma$ in Eq.~(\ref{fe}) gives the equation for $\hat{f}(y,\theta)$ to the second order in the SO interaction, which can be solved giving at $y=d$:
\begin{equation}
\partial_yn=0
\end{equation}
for the charge density and
\begin{equation}
\partial_y\mathbf{s}=2m\alpha_0\mathbf{x}\times\mathbf{s}-2m\nu eE\tau\alpha_0^2\mathbf{z}
\label{dys}
\end{equation}
for the spin density. Eq.~(\ref{dys}) is equivalent to the condition of vanishing normal (i.e., along the $y$ axis) spin-current density at $y=d$ (to the second order in $\alpha$):\cite{mishchenkoPRL04,bleibaumPRB06sd}
\begin{equation}
\mathbf{j}_a\cdot\mathbf{y}=0\,,
\label{jy}
\end{equation}
where $a=x,y,z$ label spin-density components in the Cartesian coordinates. Combining Eq.~(\ref{dys}) with the diffusion equation\cite{mishchenkoPRL04,burkovPRB04}
\begin{align}
\partial_t s_a&+\left(s_a-\delta_{ay}\nu eE\tau\alpha\right)/\tau_a=D\nabla^2s_a\nonumber\\
&+4E_F\tau\alpha\boldsymbol{\nabla}\cdot\left[\mathbf{z}\times\mathbf{a}\times\mathbf{z}\,s_z-(\mathbf{a}\cdot\mathbf{z})\,\mathbf{z}\times\mathbf{s}\times\mathbf{z}\right]
\label{de}
\end{align}
for the region with constant $\alpha$ (here, $\tau_x^{-1}=\tau_y^{-1}=\tau_z^{-1}/2=2\Delta^2\tau$) allows to solve for the spin density at $y>d$ induced by the small electric field $E$. We can, however, immediately see that since the in-plane bulk spin density $s_y$ results in the cancellation of the two terms contributing to $\partial_y\mathbf{s}$ in Eq.~(\ref{dys}), the uniform bulk solution all the way down to $y=d$ trivially satisfies both the bulk diffusion equation (\ref{de}) and the BC (\ref{dys}). Using Eq.~(\ref{sd}), we can extend the bulk solution at $y=d$ down to $y=0$, finding that $\mathbf{s}(0)=0$. The final result is thus that the nonequilibrium spin density induced by the electric field vanishes in the normal region $y\leq0$, and for $y>0$ follows the bulk value corresponding to the local Rashba parameter $\alpha(y)$, as the latter is slowly turned on.

We point out that if the term proportional to the electric field in Eq.~(\ref{dys}) is disregarded\cite{galitskiPRB06} (which could physically be justified, e.g., when the nonequilibrium spin density is induced optically at the edge in the absence of electric field), then these BC would in general describe interconversion of the $y$ and $z$ components of the spin density, which decays into the bulk on the scale of $l_s$ according to the diffusion equation (\ref{de}). This, however, gives a wrong result for the spin density generated by an electric field, as was previously noted in the case of a reflecting Hall-bar edge by Bleibaum,\cite{bleibaumPRB06bc} who obtained the same BC as Eq.~(\ref{dys}), in that case, using a very different method, and also previously postulated by Mal'shukov~\textit{et al.}\cite{malshukovPRL05sh} Finally, we note that from the structure of Eq.~(\ref{dys}), it should be clear why we calculated densities to linear order in $\alpha$ while gradients to quadratic order: The latter are governed by the spin precession in the Rashba field which is linear in $\alpha$ and the spin-generation term due to the electric field which is quadratic in $\alpha$. The second-order in $\alpha$ BC for the normal components of the density gradients thus correspond to the first-order in $\alpha$ solution of the spin-diffusion equation.

\section{Boundary Hall effect}

In this section, we offer a simple physical explanation for the spin-density jump across the interface, which will also allow us to extend our findings to more general boundary configurations. Because there is no net drift velocity in the $y$ direction, the term proportional to $\alpha^\prime$ in Eq.~(\ref{Hsoxy}) has much less effect than the term proportional to $\alpha$.  The primary role of $\alpha^\prime$ is to cause relaxation of  $s_y$, which takes place only on the very long time scale $\tau_s$ or the corresponding  length scale $l_s$. In order to understand variations on the much shorter length scale $d$, we may therefore set $\alpha^\prime=0$ in $H_{\rm SO}$. Then, $s_y$ is conserved and $\alpha({\bf{r}})$ is equivalent to a vector potential in the $x$ direction, 
\begin{equation}
\tilde{A}_{x}({\bf{r}})=\pm(m/e)\alpha({\bf{r}})\,,
\end{equation}
for electrons with $\sigma_y=\pm1$, in addition to a potential independent of $\sigma_y$, which is $\propto\alpha^2$.  Thus,  electrons with $\sigma_y=\pm1$ feel an effective orbital magnetic field in the $z$ direction, given by
\begin{equation}
\tilde{B}_z=\mathbf{z}\cdot(\boldsymbol{\nabla}\times\mathbf{\tilde{A}})=\mp(m/e)\partial_y\alpha\,.
\label{Bz}
\end{equation} 
Due to the Hall effect in $\tilde{B}_z$, an electric field in the $x$ direction leads to a chemical potential drop in the $y$ direction, which has opposite sign for the two spin states, leading to a difference of populations, and hence a jump in $s_y$, across the interval $0<y<d$. The chemical potential jump may be calculated from the standard Drude formula for the Hall resistance. Multiplying this by the density of states $\nu$, we find from Eq.~(\ref{Bz}),
\begin{equation}
s_y(y)-s_y(0)=\nu eE\int_0^ydy^\prime\tau(y^\prime)\partial_y\alpha(y^\prime)\,,
\label{syy}
\end{equation}
where $\tau(y^\prime)$ is the transport scattering time at position $y^\prime$. In the case where $\tau$ is independent of $y$, the right-hand side of Eq.~(\ref{syy}) for $y>d$ is just $\nu eE\tau(\alpha_0-\alpha_1)$, which is precisely the difference in the bulk polarizations in the two media, far from the boundary between them. The solution of the coupled spin-diffusion equation (\ref{de}) is then simple: The polarization $s_y$ has one constant value in the region $y<0$ and another in the region $y>d$, with an effective discontinuity at $y\approx0$. There are no gradients of $s_y$ on the scale of the spin-diffusion length on either side of the boundary, and hence the diffusion equation leads to no polarization out of the plane. This is true, in principle, even if there is a difference in the electron density on the two sides of the interface, provided that the electron mobility, and hence $\tau$, is constant throughout.

In general, however, density differences caused by charging a gate over one half of the system will lead to variations in $\tau$ and hence a gradient of $\tau$ in the transition region $0<y<d$. This will cause the discontinuity in $s_y$ given by Eq.~(\ref{syy}) to deviate from the difference in bulk polarizations, forcing the existence of gradients in $s_y$ and nonzero values of $s_z$, in a region of the size of the spin-diffusion length, on either side of $y=0$. Of particular interest is the case where $\alpha\approx0$ on one side of the boundary, and the spin-diffusion length is especially long in that region. Then, if there is a gradient of electron mobility in the transition region, there can be a nonzero value of $s_z$ which extends far into the ``normal region" where $\alpha\approx0$. The value, however, will depend on the position dependence of $\tau$ as well as the behavior of $\alpha$ in the transition region.

\section{Conclusions and discussions}

We have assumed so far that the scale $d$ for variation of $\alpha$ is larger than the electron mean free path. In the case where $\tau$ is a constant, however, our results should apply even to steplike variations in $\alpha$, as long as we are not interested in the oscillating structure on the scale of the Fermi wavelength $\lambda_F$.\cite{reynosoPRB06}

In the case of a perfectly {\em{reflecting}} boundary at $y=0$, with current flow parallel to the boundary, the existence of an effective magnetic field $\tilde{B}_z$ very close to the boundary would have no physical effect: The generated Hall voltage is added to a potential which is in any case infinite on the insulating side of the interface. The solution of the diffusion equation (\ref{de}) for constant $\alpha$ at $y>d$ with BC (\ref{dys}) gives a spin polarization $s_y$ which is constant and the same as in the bulk, for points further than $d$ from the boundary. We find that $s_z=0$ in this case, which is true even when $d\to0$,  in agreement with Refs.~\onlinecite{malshukovPRL05sh} and \onlinecite{bleibaumPRB06bc}.

Finally, we note that in the case where the electric field is normal to a transmitting boundary, there should be no discontinuity in the spin polarization across the interface. If $\alpha$ had depended on $x$ rather than $y$, in the discussion above, the vector potential $\tilde{A}_x$ would have been purely longitudinal, giving no physical effect.

In summary, we have developed a method, starting from the quantum kinetic equation, for studying spin Hall effects in a system with a position-dependent Rashba coupling constant $\alpha$, in the ``dirty limit" $\Delta\tau\ll1$. We considered systems where $\alpha$ depends only on $y$ and changes discontinuously on the scale of the spin-diffusion length $l_s$, from one constant value to another, with a uniform applied electric field $E$ in the $x$ direction, and we derived boundary conditions for the coupled spin and charge diffusion equations, which apply away from the discontinuity. In general, we find a discontinuity in the $y$ component of the spin density $\bf{s}(\bf{r})$ at the boundary. In the case where 
the transport scattering time $\tau$ is independent of position, this leads to values of ${\bf{s}}$ away from the boundary that are the same as would occur for a uniform system with the local value of $\alpha$, namely, $s_y=\nu eE\tau\alpha$ and $s_x=s_z=0$. Thus, in contrast to Ref.~\onlinecite{galitskiPRB06}, we find no lateral spin injection into a normal conductor and no out-of-plane spin density, when the electron mobility is uniform. The related spin Hall current extraction proposal of Ref.~\onlinecite{adagideliPRL05} can therefore only work under the assumption of an inhomogeneous mobility close to the interface. We also find $s_z=0$ at a {\em reflecting} boundary, in contrast to Ref.~\onlinecite{galitskiPRB06}, but in agreement with Refs.~\onlinecite{malshukovPRL05sh} and \onlinecite{bleibaumPRB06bc}. On the other hand, for a transmitting boundary, if $\tau$ varies across the transition region, one can get spin injection and $s_z\neq0$.

\acknowledgments

We are grateful to I.~Adagideli, L.~Balents, G.~E.~W. Bauer, A.~A. Burkov, V.~M. Galitski, and E.~I. Rashba for valuable discussions. This work was supported in part by the Harvard Society of Fellows, NSF Grants Nos. PHY 01-17795 and DMR 05-41988, and the Research Council of Norway through Grant Nos. 158518/143, 158547/431, and 167498/V30.

\end{document}